# MRI/TRUS data fusion for prostate brachytherapy. Preliminary results.

Christophe Reynier, Jocelyne Troccaz
*TIMC Laboratory, School of Medicine, F-38706 La Tronche cedex, France*

Philippe Fourneret[1], André Dusserre[1], Cécile Gay-Jeune[3], Jean-Luc Descotes[2], Michel Bolla[1], Jean-Yves Giraud[1]
[1]*Radiotherapy department*, [2]*Urology and renal transplant department*, [3]*Radiology and MRI departments*
*CHU de Grenoble, BP 217,*
*F-38043 Grenoble cedex 09, France*

Prostate brachytherapy involves implanting radioactive seeds ($I^{125}$ for instance) permanently in the gland for the treatment of localized prostate cancers e.g. cT1c-T2a N0 M0 with good prognostic factors. Treatment planning and seed implanting are most often based on the intensive use of transrectal ultrasound (TRUS) imaging. This is not easy because prostate visualization is difficult in this imaging modality particularly as regards the apex of the gland and from an intra- and inter-observer variability standpoint. Radioactive seeds are implanted inside open interventional MR machines in some centres. Since MRI was shown to be sensitive and specific for prostate imaging whilst open MR is prohibitive for most centres and makes surgical procedures very complex, this work suggests bringing the MR virtually in the operating room with MRI/TRUS data fusion. This involves providing the physician with bi-modality images (TRUS plus MRI) intended to improve treatment planning from the data registration stage. The paper describes the method developed and implemented in the PROCUR system. Results are reported for a phantom and first series of patients. Phantom experiments helped characterize the accuracy of the process. Patient experiments have shown that using MRI data linked with TRUS data improves TRUS image segmentation especially regarding the apex and base of the prostate. This may significantly modify prostate volume definition and have an impact on treatment planning.

## I. INTRODUCTION

Prostate cancer is the most common cause of cancer in men in many countries including the European Community and NorthAmerica. 40209 cases were reported out of a 60 million population in France in 2002 [1] matching a standardized incidence of 75.3 (rate per 100000) and 10004 individuals subsequently died. Many options – careful watch, conventional or laparoscopic surgery, brachytherapy, conventional or conformal external radiotherapy including IMRT, focused ultrasound, cryosurgery, etc. – are available and based on a multi-disciplinary approach with informed patient consents. Individual screening from blood analysis using PSA allows for increased detection of early and localized stages. Recent progresses tend towards improving local control whilst minimizing side effects.

Using permanently placed radioactive seeds (a technique known as brachytherapy) back in the 1970s produced initial results that were as good as those obtained with other radiotherapy techniques and presented low impotence risks or other side effects. However seeds were placed freehand during laparotomy at that time [2] and resulted in non-homogeneous irradiation where cold regions may have contributed to disappointing clinical results.

Treatment planning and seed implantation is now based on the intensive use of transrectal ultrasound imaging. A stepper is used to acquire parallel transverse images with constant inter-slice distance. Moreover, needle insertion through the perinea is made easier via a grid, also called template, rigidly connected to the TRUS probe. This improved clinical outcome but such an approach is still limited by two factors. Firstly, the prostate apex and base are sometimes difficult to visualize in the transverse TRUS images and, secondly, intra- and inter-observers variations are very frequent in ultrasound image interpretation. These problems could possibly be solved with image processing automation. Results regarding prostate automatic segmentation [3] whilst encouraging are generally obtained for images of the prostate acquired rather far from its extremities and therefore easier to process.

Whilst prostate imaging is still controversial, Magnetic Resonance Imaging (MRI) has a great potential. Its sensitivity and specificity have been demonstrated. Radioactive seeds are currently implanted inside an open interventional MR (iMRI) machine in a small number of centres [4]. Robots may also be used in conjunction with CT or MRI imaging for improved accuracy and increased possible trajectories [5]. Yet, iMRI is prohibitive for most centres and makes surgical procedures much more complex. Moreover, even when using iMRI for prostate procedures (biopsy or brachytherapy), pre-operative MRI is also acquired because it provides enhanced image quality and is correlated with iMRI [6].

The alternative solution suggested is "bringing the MR *virtually*" in the operating room using MRI/TRUS data fusion. This approach allows refining prostate localization in the TRUS images with MRI data acquired pre-operatively. A similar approach was developed for biopsy guidance by Kaplan et al [7]. Those authors suggested registering MRI to TRUS by matching six fiducial prostate points defined in both modalities. Whereas, for sake of accuracy and robustness improvement in data fusion, we elected to record large sets of 3D points. In the presented approach, contours segmented in the intra-operative TRUS data with a satisfactory level of confidence are used for registration. MRI data are then superimposed to TRUS images to verify and improve TRUS segmentation at the apex and base of the prostate in particular. The objective is to improve data acquisition, homogeneity of dosimetry with a better definition of the D90 isodose and the reliability of urethra and rectum Dose Volume Histograms (DVH). This might help control cancer while reducing bladder, rectum and urethra side effects.

## II. MATERIAL AND METHODS

Two types of protocols are used for prostate brachytherapy. The first is based on a pre-planning stage whilst the treatment planning stage is taking place intra-operatively in the second one. The second protocol is that applied in our institution.

### A. Proposed method

The PROCUR system presented in this paper is technically approached as follows. An MRI exam is performed with a transrectal coil and acquisition proceeds in the 3 orthogonal phases on the day preceding seed implantation. The patient is lying supine. Data are recorded in the DICOM format and transmitted to the computer for image processing. The contours of the prostate are manually segmented in the three planes. Data are segmented from the sagittal and coronal planes and superimposed onto those contours to facilitate apex segmentation in the MRI horizontal images. This step helps construct a three dimensional (3-D) cloud of *n* points, $S^{MRI}$, representing the prostate in the MRI reference frame $R_{MRI}$.

Intra-operatively, the patient is placed in the gynaecologic position. The TRUS probe is inserted and transverse images are collected regularly with a stepper as in the conventional procedure. The prostate is swept as a whole and approximatively 8 to 10 parallel images are collected. The urologist delineates the prostate on every image whenever possible. The data are then fed in the computer and the urologist segmentation is recorded. This results in a second set of $m$ 3-D points, $S^{TRUS}$, representing the prostate in the ultrasound reference system $R_{TRUS}$.

The next stage involves registering both cloud points automatically. The system computes the $T_{TRUS/MRI}$ transform between $R_{TRUS}$ and $R_{MRI}$ allowing for optimal $S^{TRUS}$ to $S^{MRI}$ superimposition. This is in fact minimizing the following energy as a function of the $T_{TRUS/MRI}$ unknown:

$$E(T_{TRUS/MRI}) = \sum_{i=1}^{m} \frac{1}{\sigma_i^2} dist(S^{MRI}, T_{TRUS/MRI}(M_i^{TRUS}))^2$$

where $M_i^{TRUS}$ is the $i^{th}$ point belonging to $S^{TRUS}$, $\sigma_i$ is the estimated data-related error and $S^{MRI}$ represents the MRI data set. $dist(S,P)$ is the Haussdorf distance between a point $P$ and a surface $S$.

A pre-registration consisting in superimposing the $S^{TRUS}$ and $S^{MRI}$ centres of gravity initializes $T_{TRUS/MRI}$ before minimization. Two types of registration have been tested. In the first one called rigid registration, $T_{TRUS/MRI}$ is a homogeneous transform combining rotations and translations (6 parameters). The second one is an elastic registration where $T_{TRUS/MRI}$ is much more complex allowing for rotation and translation between data sets as well as local deformations or distortions. Those methods are derived from the octree-spline elastic registration published by Szeliski and Lavallée [8]. The optimization procedure is the Levenberg-Marquardt algorithm (see [9]). Elastic registration makes use of an adaptive, hierarchical and regularized free-form deformation of one volume to the other coordinate system. Since $S^{MRI}$ density is much higher than $S^{TRUS}$ ($n>>m$), the octree-spline construction is based on $S^{MRI}$. The result is a 3-D function $f_{TRUS/MRI}$ transforming any point given in $R_{TRUS}$ to $R_{MRI}$.

Once the data have been registered (see figure 1) the software may compute the corresponding slice in the MRI transverse volume for every acquired TRUS image using $f_{TRUS/MRI}$ and the interpolation tools. An image presenting both the TRUS and MRI data is computed to help the urologist visualize results and verify or improve prostate segmentation. The resulting image is split into four quadrants with a virtual cross: the upper right and lower left quadrants display TRUS data whilst the upper left and bottom right ones present the corresponding re-sliced MRI data (see figure 2). The users are allowed to move the cross in this bi-modality image. This representation helps them explore prostate limits in both modalities. Contours extracted from the MRI data may also be superimposed to the TRUS images. MRI data help confirm or modify the original segmentation close to the apex and base of the prostate in particular. A 3-D representation is also available to the clinician. Three intersecting orthogonal MRI slices are computed in the three MRI volumes for any given needle i.e. a point defined in a transverse TRUS image at a normal direction to the image plane.

This improved segmentation can then be used for treatment planning. The 3-D view may facilitate appreciating the needle position with respect to the whole gland and surrounding structures.

**B.     Materials**

The MRI system is a Philips Gyroscan ACS II 1.5 Tesla operated with a transrectal coil. T2 Turbo Spin Echo sequences are used for transverse, sagital and coronal acquisitions. The repetition time is 1800ms and echo time 120ms. The region explored is 15cmx15cm; image resolution is 256x256. The slice thickness is 3mm. The echographic system is a BK-Medical one. The stepper is designed to acquire TRUS images every 5mm. The probe is a *8551* model (5-10Mhz intrarectal probe). The TRUS image size is 576x768 pixels. A tissue-equivalent ultrasound prostate phantom[1] from CIRS (Computerized Imaging Reference Systems, Inc.) was used for technical evaluation and accuracy quantification. The PROCUR system is written in Visual C++. The computer is a PC Intel Pentium4, 1.6GHz with 256MB of RAM and equipped with a Matrox MeteorII video frame grabber. The PC is operated with Windows2000 Professional.

**C.     Validation approach**

It must first be stated that treatment was planned and administered as usual to the patients participating in those evaluations. Imaging data (pre-operative MRI and intra-operative TRUS) were recorded. The PROCUR system was evaluated post-operatively from acquired data. Some evaluations were also performed with the phantom.

*1. Technical evaluation*

Registration is a key step in the PROCUR system. The registration methodology as presented in section II.1 involves minimizing the distance between two sets of 3-D points. It can be evaluated in several ways. Firstly, one can visually confirm that the two sets of points have a good fit. This evaluation is purely qualitative but should registration be poor, it may be useful to determine where mismatches were made. A second and more interesting qualitative computed transform evaluation is based on user interaction with the bi-modality image produced after data fusion. The continuity of prostate contours from one modality to the other is a very good indicator of registration quality.

Quantitatively, information regarding the distance remaining between both sets of points is an indicator of registration quality. Mean, maximum, minimum values as well as standard deviations are reported in section III.1. However, small residual distances are not true equivalent of optimal registration and optimization algorithms may be trapped into local minima of the energy function. Pre-registration reduces this risk but it is not completely eliminated. A second registration accuracy assessment methodology consequently needs to be defined without any gold standard – the exact computed transform value cannot be determined – and it should be as minimally invasive as possible for the patients – i.e. avoiding marker placement as in [10] for instance. It was decided to verify the registration accuracy of the urethra. This is a significant structure in treatment planning; it is internal to the prostate and is not included in registration data. As far as the phantom is concerned, the urethra is easily visible in both modalities. It has to be visible for patients and this is why the procedure described below was adopted. A probe is inserted into the patient's urethra and retained for the MRI examination as well as the implantation procedure the next day. This allows for urethra visibility in both MRI data and TRUS images. The PROCUR system is then operated as described in section II.A and produces a $f_{TRUS/MRI}$ value. The bi-modality image is computed for each echographic plane. The urethra position (lumen centre) is recorded in the TRUS and MRI components for every bi-modality image (see figure 3). The distance between those two points is computed. Mean, minimum, maximum values and standard deviations valid for phantom experiments and series of 4 patients are presented in section III.1

---

[1] Model 053 (see http:\\www.cirs.com)

## *2. Potential impact on treatment*
Other evaluations of a more clinically advanced level have also been performed. The effects resulting from the use of three orthogonal MRI acquisitions have been quantified for series of 8 patients. The objective was to compare MRI transverse image segmentation with and without the assistance of the two other volumes. The effects of data fusion and MRI superimposition on prostate delineation were determined in two different manners. First, the number of contoured echographic images with and without data fusion have been compared close to the apex and to base of the gland for 11 patients. Secondly the prostate volume was computed without and with data fusion for the same patients. These data have a serious potential impact for the treatment planning stage. Results are presented in section III.2.

A dose volume histogram (DVH) was computed with the MRI-enhanced echographic data to predict the potential impact of volume modification on treatment outcome. It was compared to that applicable for the conventional procedure. A comparison is presented for one patient (No 5).

## III. RESULTS
### III.1 Technical evaluation
**Phantom**
11 images of the phantom were obtained (see figures 4 and 5) with the TRUS probe (5.2MHz). Registration results are presented in table I. One can see a satisfactory match for prostate contours in registered modalities (Fig. 5).

| **Table I** | *Residual distance error between $S^{TRUS}$ and $S^{MRI}$ (mm)* | |
|---|---|---|
| *Registration method* | Mean [min; max] | Std. Dev. |
| Rigid | 1.62 [0.34; 2.64] | 0.40 |
| Elastic | 1.07 [0.01; 2.75] | 0.41 |

The urethra match measurements on bi-modality images after registration are presented in the following table. These values correspond to the 11 bi-modality computed images of the phantom. Please note that the phantom urethra diameter is about 10mm.

| **Table II** | *Distance between the urethra detected on TRUS and on MRI (mm)* | |
|---|---|---|
| *Registration method* | Mean [min; max] | Std. Dev. |
| Rigid | 1.30 [0.58; 2.63] | 0.58 |
| Elastic | 1.57 [0.82; 2.87] | 0.62 |

Rigid and elastic registrations are very similar because the phantom is non-deformable.

**Patients**
Table III reports results obtained for 11 patients.

| **Table III** | *Residual distance of the $S^{TRUS}$ and $S^{MRI}$ surface points* | |
|---|---|---|
| *Patient* | Rigid registration Mean [min; max] ± Std. Dev. | Elastic registration Mean [min; max] ± Std. Dev. |
| 1 | 1.48 [0.84; 2.12] ± 0.27 | 1.13 [0.09; 4.03] ± 0.54 |
| 2 | 1.58 [0.63; 2.59] ± 0.39 | 1.26 [0.10; 5.68] ± 0.76 |
| 3 | 1.33 [0.73; 1.93] ± 0.23 | 0.95 [0.07; 2.79] ± 0.46 |
| 4 | 1.29 [0.50; 2.28] ± 0.30 | 1.08 [0.13; 4.00] ± 0.53 |
| 5 | 1.31 [0.43; 2.24] ± 0.37 | 1.03 [0.07; 6.51] ± 0.59 |
| 6 | 1.30 [0.73; 1.80] ± 0.22 | 1.09 [0.15; 3.08] ± 0.44 |
| 7 | 1.38 [0.72; 2.09] ± 0.25 | 1.09 [0.06; 2.48] ± 0.40 |
| 8 | 1.25 [0.65; 1.81] ± 0.25 | 1.09 [0.04; 3.25] ± 0.46 |
| 9 | 1.32 [0.44; 2.25] ± 0.33 | 1.21 [0.09; 4.91] ± 0.57 |
| 10 | 1.25 [0.42; 2.29] ± 0.37 | 1.07 [0.11; 2.60] ± 0.44 |
| 11 | 1.42 [0.19; 3.22] ± 0.51 | 1.17 [0.09; 5.20] ± 0.76 |
| *Average patient* | *1.36 [0.57; 2.24] ±0.22* | *1.11 [0.09; 4.05] ±0.54* |

The verification on urethra (lumen centre) is reported in table IV for 4 out of these 11 patients. The diameter of the urethra is 5 mm approximately.

| **Table IV** | *Distance between the urethra detected on TRUS and on MRI (mm)* | |
|---|---|---|
| *Patient* | Rigid registration<br>Mean [min; max] ± Std. Dev. | Elastic registration<br>Mean [min; max] ± Std. Dev. |
| 5 | 3.68 [1.39; 6.17] ± 1.89 | 2.05 [0.31; 3.77] ± 1.50 |
| 6 | 3.52 [0.46; 9.24] ± 3.24 | 2.96 [0.46; 8.25] ± 3.06 |
| 7 | 3.53 [0.49; 3.53] ± 1.00 | 1.51 [0.46; 2.19] ± 0.67 |
| 9 | 2.03 [0.16; 5.71] ± 1.67 | 1.73 [0.56; 3.70] ± 1.04 |
| *Average patient* | *2.90 [0.62; 6.16] ±1.95* | *2.07 [0.45; 4.48] ±1.57* |

In Figure 6, the urethra distance is presented as a function of the slice number. Each curve corresponds to one patient. It clearly appears that errors are systematically higher for apical images and decrease fairly regularly. This will be discussed in section IV. Figure 7 shows the average urethra distance obtained for each slice and all 4 patients.

### *III.2 Pre-clinical evaluation*
**MRI segmentation**

Sagital and coronal MRI acquisitions helped increase the number of segmented slices in the MRI horizontal plane (see table V). On average, 1.12 slices representing 9.78% could be segmented additionally for the prostate base. 1.62 slice representing 14.13% were added for the apex.

| **Table V** | *Contribution of the sagittal and coronal MRI data to the segmentation of the prostate on the transverse MRI volume* | | |
|---|---|---|---|
| *Patient number* | *Nb of segmented transverse slices* | *Nb of additional slices (base)* | *Nb of additional slices (apex)* |
| 1 | 10 | 2 | 1 |
| 2 | 10 | 1 | 0 |
| 3 | 14 | 1 | 2 |
| 4 | 11 | 1 | 2 |
| 5 | 14 | 0 | 3 |
| 6 | 13 | 1 | 2 |
| 7 | 10 | 1 | 2 |
| 8 | 10 | 2 | 1 |
| *Mean* | *11.5* | *1.125* | *1.625* |
| *%* | *-* | *9.78%* | *14.13%* |

**TRUS segmentation**

Using MRI data resulted in a modification of the number of segmented TRUS transverse images (see table VI): slices were added (or removed) for 4 (respectively 1) of the 11 patients i.e. 36.16% (or 9.1% respectively) of the cases. The base segmentation was modified for 5 of

Page 6

the 11 patients i.e. 45.45% of the cases. The apex segmentation was modified for 4 of the 11 patients i.e. 36.36% of the cases. Figure 8 shows a case where the prostate apex was not segmented on TRUS; MRI data provided additional information and enabled contour definition.

| Table VI | | Contribution of MRI data to the segmentation of apex and base of the prostate on TRUS images | | | |
|---|---|---|---|---|---|
| Patient number | Before registration | After registration (number of images added or removed) | | | |
| | | Apex | | Base | |
| | | Rigid | Elastic | Rigid | Elastic |
| 1 | 11 | 0 | 0 | 0 | 0 |
| 2 | 12 | -1 | 0 | -2 | -1 |
| 3 | 8 | 0 | 0 | 0 | 0 |
| 4 | 8 | 0 | 0 | 0 | 0 |
| 5 | 7 | 0 | 0 | 0 | 0 |
| 6 | 8 | 0 | 0 | 0 | 0 |
| 7 | 8 | 0 | 0 | 1 | 1 |
| 8 | 5 | 1 | 1 | 1 | 1 |
| 9 | 10 | 0 | 0 | 0 | 0 |
| 10 | 7 | 0 | 1 | 1 | 1 |
| 11 | 9 | 2 | 1 | 2 | 1 |
| Modified segmentation | - | 27.27% of the cases | 27.27% | 45.45% | 45.45% |
| | | 36.36% | | 45.45% | |

**Volume measurement**

Prostate volume was measured for the original TRUS segmentation and modified segmentation resulting from MRI enhancement in each patient. The following formula was applied to compute this volume:

$$V = \sum_{i=1}^{n-1} \frac{S_i + S_{i+1} + \sqrt{S_i * S_{i+1}}}{3} * d$$

Where $n$ is the number of slices, $S_i$ the surface of the $i^{th}$ slice and $d$ the inter-slice distance.

As regards the phantom, the volume corresponding to the original segmentation is 65.92cc; it is 67.97cc after elastic registration and MRI-enhancement. This corresponds to a difference of 2.05cc and is equivalent to 3.11% of the original volume. It was not possible to compare these computed volumes with the real one because the dimensions given by the phantom provider are averages only. The difference between those two values mainly comes from the significant increase in surface (5.94cm$^2$) of the first slice (apex) for which contour detection on the original TRUS image proved very difficult. In absolute (or signed) values, the surface difference (cf. figure 9) is: max = 5.94cm$^2$, min = 0.05cm$^2$, mean = 1.01cm$^2$, std.dev. = 1.69cm$^2$ (max = 5.94cm$^2$, min = -0.66cm$^2$, mean = 0.63cm$^2$, std.dev. = 1.88cm$^2$ respectively).

As can be seen on figure 10, the volume computed from enhanced images after elastic registration is higher than the original volume for all patients. In terms of percentage variation relatively to the initial volume, the minimum, maximum, mean values and the standard deviations are: min = 1.03% (patient No 6), max = 48.24% (patient No 8), mean =15.86%, std. dev. = 13.57%. The largest three volume differences (patients No 7, 8 and 10) correspond to cases where one additional echographic image was segmented after registration (see table VI). Concerning the other two patients (No 2 and 11) for whom the number of echographic images changed, the volume difference is lower. As regards patient No 2, the removed slice is compensated in terms of volume by an increase in computed surfaces on the other slices. As regards patient No 11, the added slice is partly compensated by the fact that the largest surface (slice No 5) is smaller in modified data compared to the original ones.

**Dose volume histograms**
Planned treatment in clinical practice is such that 90% of prostate volume should receive between 160 and 180Gy. The D90 planned on the original segmented contours of patient No 5 corresponds to 165Gy i.e. 90% of the prostate is supposed to receive at least 165Gy. When the D90 is recomputed from modified contours after elastic registration, the corresponding value is 105Gy (i.e. 90% of the prostate is supposed to receive at least 105Gy). 68.45% only of the prostate would receive at least 165Gy for these modified contours. The difference between those two values is very significant (-36%). However, this is an individual test from which no conclusion can be drawn. Still, as pointed out in [11], image processing may significantly contribute to inaccuracy of dose distribution and measurement. Our results are very preliminary but they show that a significant difference in TRUS prostate definition coming from MRI enhancement may have a very large impact on potential clinical outcomes.

### IV. DISCUSSION AND CONCLUSION

Concerning surface registration, qualitative results based on visual evaluation are generally quite positively appreciated from a clinician's standpoint. Quantitative results are close to 1mm in average and this is very satisfactory as the different sources of errors (image resolution, manual segmentation[2]) are taken into consideration. Rigid and elastic registrations give rather similar results in terms of residual distance measurements for both sets of points. It is likely that the rather similar positions of the patient during MRI and TRUS acquisition, which both involve placing an intra-rectal sensor, result in rather similar deformations of the prostate, if any. Prostate motion depending on bladder and rectum filling are ideally taken into account with rigid registration. Elastic registration produces lower mean and minimum values as regards the residual distance between data, but the standard deviation and maximum residual distance are somewhat higher. This can be explained by the fact that elastic registration imposes regularization to retain surface smoothness. This means that reducing the distance of one TRUS point closer to the MRI surface may increase the distance of an adjacent TRUS point to retain the TRUS surface continuity. Elastic registration will be the preferred methodology because it may render potential deformations. Both processes are real-time ones.

Our method differs from [7] on several accounts: the density of registered data (6 points versus large data sets) and the type of registration (point-to-point versus set-to-set and rigid versus rigid or elastic). The accuracy of small data sets point-to-point matching is highly dependent on the ability to precisely define the corresponding points in both modalities. However, the approach presented in [7] helps simplify the protocol. In the PROCUR system, the fact that contours are used without any need for explicit point-to-point equivalence between MRI and TRUS data has a related cost i.e. MRI prostate segmentation (TRUS segmentation is a standard stage in brachytherapy protocols). No quantitative comparison can be made with our work since no quantitative results are provided in reference [7].

As regards urethra experiments, figures 6 and 7 clearly show that measurements are dependent on the TRUS slice position for the patients. The distance measured between both modalities is always higher close to the apex. This may be partly explained by the fact that the

---

[2] In MRI images where pixel size is 0.58mm x 0.58mm for instance, a variation of 2 pixels in manual segmentation is equivalent to a 1.16mm distance on the contour ..

Page 8

TRUS apex slices often are the smallest surfaces and the most difficult to read. The uncertainty regarding urethra position is certainly a factor contributing to this error. Moreover, defining the lumen centre manually may be somewhat difficult and perhaps inaccurate even for a 'satisfactory' TRUS image. Errors are much smaller with the phantom because the urethra boundary is much more visible on TRUS images. A semi-automated procedure would probably reduce the centre localization inaccuracy. Those results do not seem to be correlated to the quality of surface registration but this would have to be confirmed with statistical analysis for a larger number of cases. Reference [12] describes CT/MRI registration for dose evaluation using urethral catheters. The authors underline that the least precise results were obtained near the prostate base and this is explained by potential balloon buoyancy. Concerning our experiments, the potential variability of urethral probe traction between MRI acquisition and the intra-operative situation could only be a low contributor to those results. One can also observe that our results globally improved from patient to patient and that may have been a learning curve effect for the placement of the urethral probe in a stable position. This should be confirmed with a significant number of cases.

We demonstrated with series of 8 patients that using three MRI volumes acquisition facilitates prostate segmentation for the apical region and the base in particular. We also demonstrated that MRI enhancement allows modifying the number of segmented TRUS images especially at the apex and base of the prostate. Experiments made regarding volume measurements have shown that MRI enhancement always results in computed volume increases. The formula used for volume computation may be discussed and different expressions could be used (a simple average for two successive slices or a formula taking into account the fact that a prostate section may be unreported before the first slice and after the last one because of TRUS step or MRI inter-slice distance – both were experimented). However, general conclusions can be drawn because the formulae used for TRUS and MRI-enhanced images are identical. We have seen that underestimating the prostate volume in the TRUS data could have a very significant potential impact on dose distribution. This needs to be confirmed with a clinical evaluation intended to correlate clinical factors (post-brachytherapy PSA measurements for instance) with those simulations.

The next project stage is the clinical evaluation. Other long-term extensions could also be envisioned. Introducing MRI affords extensive opportunities. Several centres are attempting to evaluate its potential use in cancer localization (see reference [13] for instance) and this would allow for more precise and selective seeds positioning. In the same vein, seed placement could be based on MRI navigation with pre-operative data as brain and spine surgeons have been doing for many years now [14,15].

**ACKNOWLEDGMENTS**
This project is supported by the French Ministry of Health (Projet Hospitalier de Recherche Clinique PHRC "Prostate-écho").

**List of figures:**

Figure 1: *Data superimposition after registration*

Figure 2: *Data fusion*: the computed image presents MRI (top-left/bottom-right) data and TRUS data (top-right/bottom-left) for a given TRUS acquisition plane. Two examples.

Figure 3: *Urethra distance computation*

Figure 4: *Prostate TRUS phantom:* (left) photograph (right) inside.

Figure 5: *Phantom experiment*: (a) MRI image – (b) TRUS image – (c) composite image after data fusion

Figure 6: *Urethra distance with elastic registration.* Each curve corresponds to one patient or the phantom.

Figure 7: *Mean residual distance of the urethra*

Figure 8: *Segmentation of the apical region*: (a) original non segmented TRUS image – (b) corresponding MRI recomputed slice – (c) contours from the MRI volumes (blue for frontal, yellow for sagital and red for transverse)

Figure 9: *Surface analysis (phantom)*: elastic registration.

Figure 10: *Volume analysis (patients)*: elastic registration case

**List of tables**

Table I: *Residual distance error between $S^{TRUS}$ and $S^{MRI}$ (mm)*. Phantom experiments.

Table II: *Distance between the urethra detected on TRUS and MRI (mm)*. Phantom experiments.

Table III: *Residual distance of the $S^{TRUS}$ and $S^{MRI}$ surface points*. Patient experiments.

Table IV: *Distance between the urethra detected on TRUS and MRI (mm)*. Patient experiments.

Table V: *Contribution of sagittal and coronal MRI data to segmentation of the prostate on the transverse MRI volume*

Table VI: *Contribution of MRI data to the segmentation of apex and base of the prostate on TRUS images.*

Figures

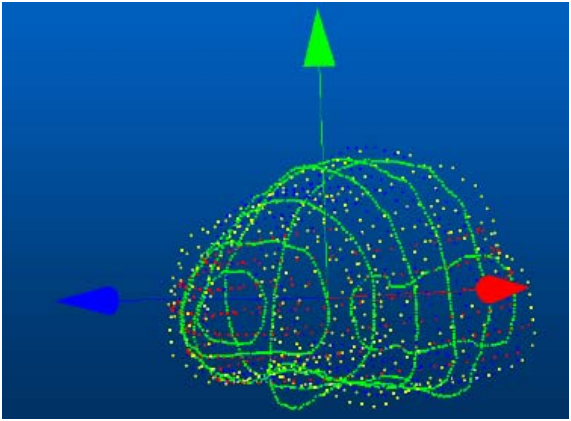

Figure 1: *Data superimposition after registration*

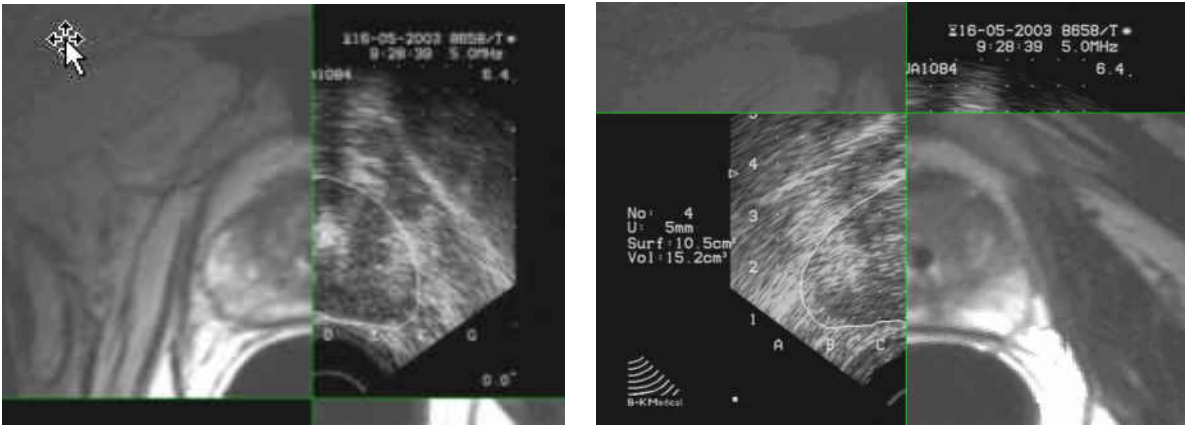

Figure 2: *Data fusion*: the computed image presents MRI (top-left/bottom-right) data and TRUS data (top-right/bottom-left) for a given TRUS acquisition plane. Two examples.

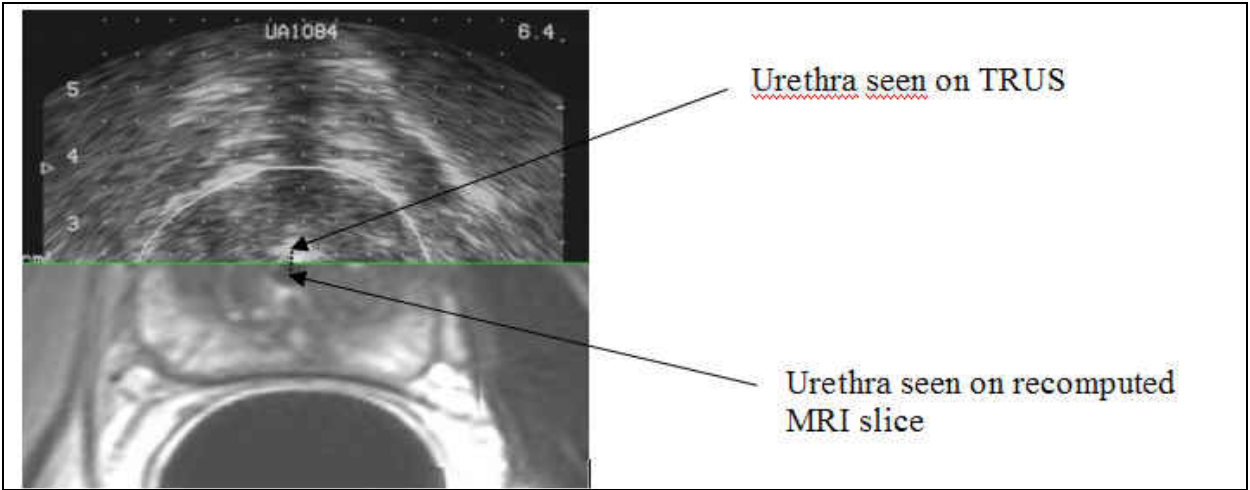

Figure 3: *Urethra distance computation*

Page 13

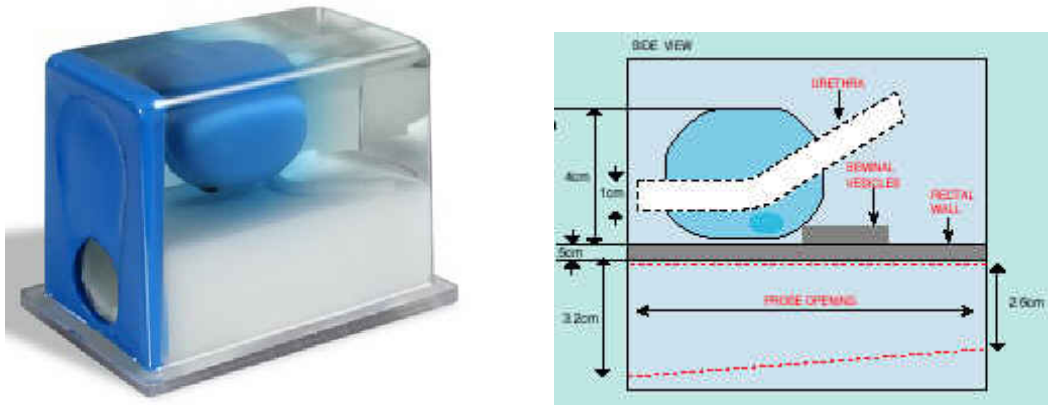

Figure 4: *Prostate TRUS phantom:* (left) photograph (right) inside.

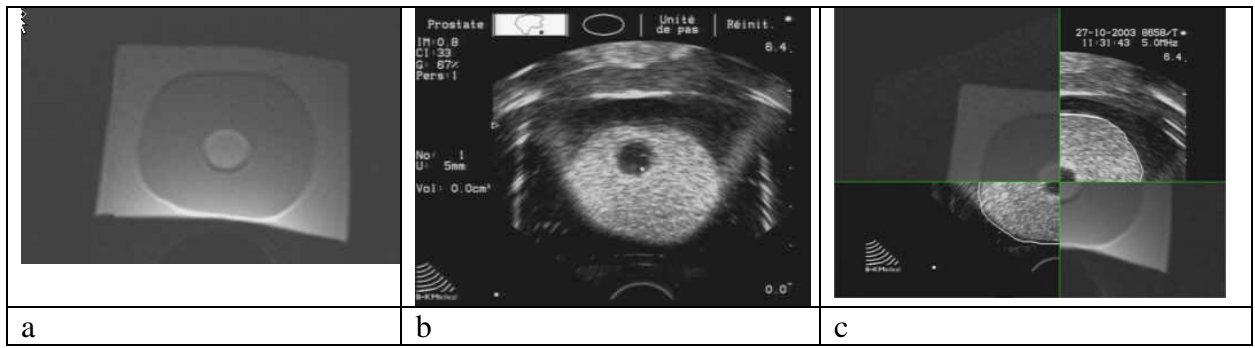

| a | b | c |

Figure 5: *Phantom experiment*: (a) MRI image – (b) TRUS image – (c) composite image after data fusion

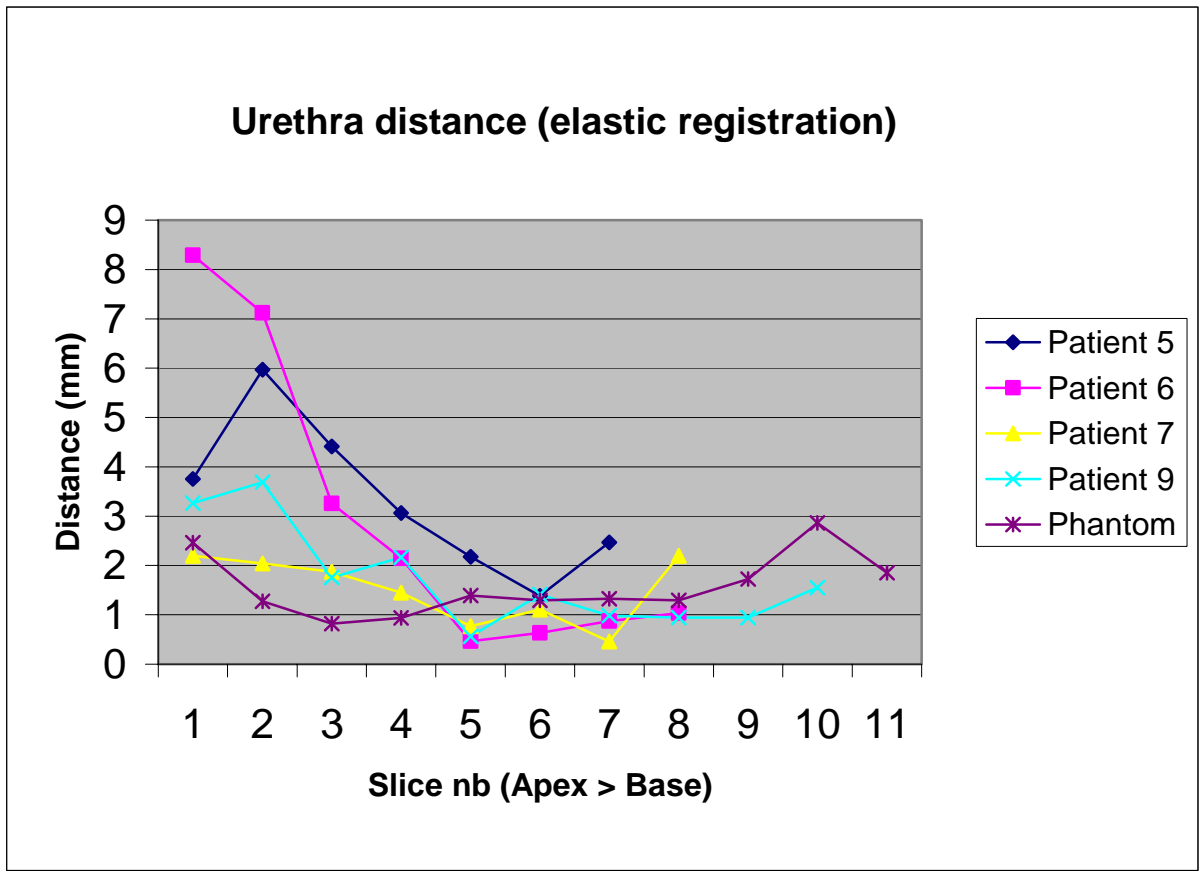

Figure 6: *Urethra distance with elastic registration.* Each curve corresponds to one patient or the phantom.

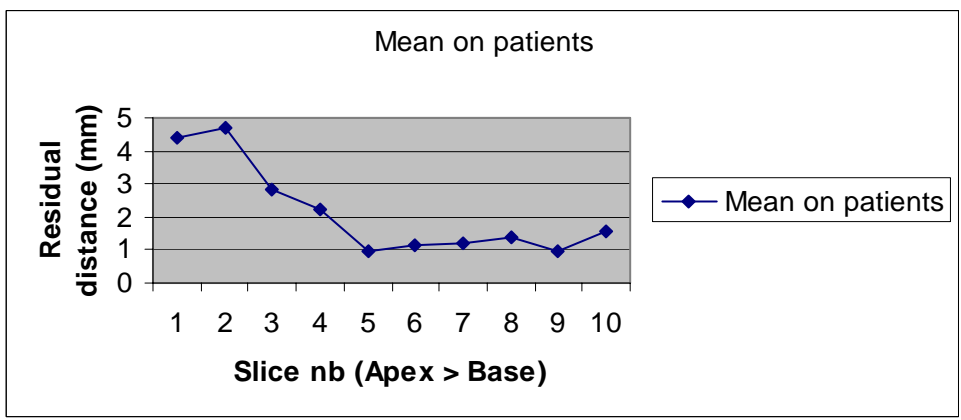

Figure 7: *Mean residual distance of the urethra*

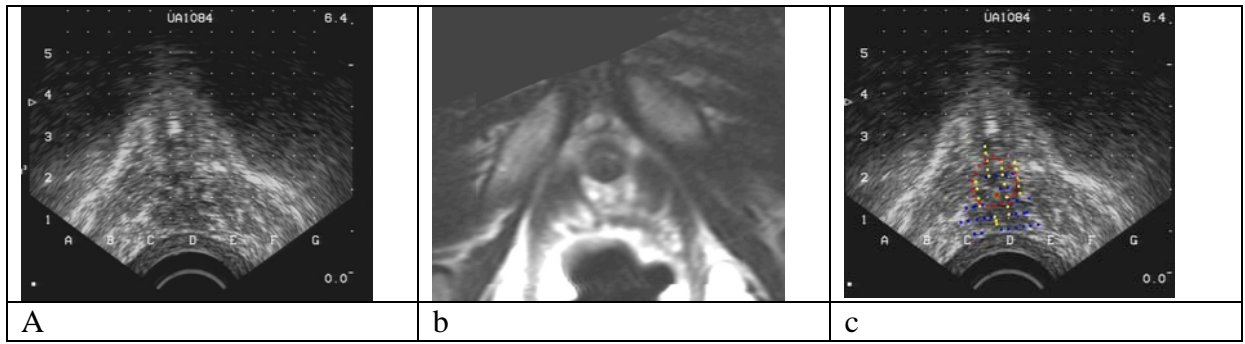

| A | b | c |

Figure 8: *Segmentation of the apical region*: (a) original non segmented TRUS image – (b) corresponding MRI recomputed slice – (c) contours from the MRI volumes (blue for frontal, yellow for sagittal and red for transverse)

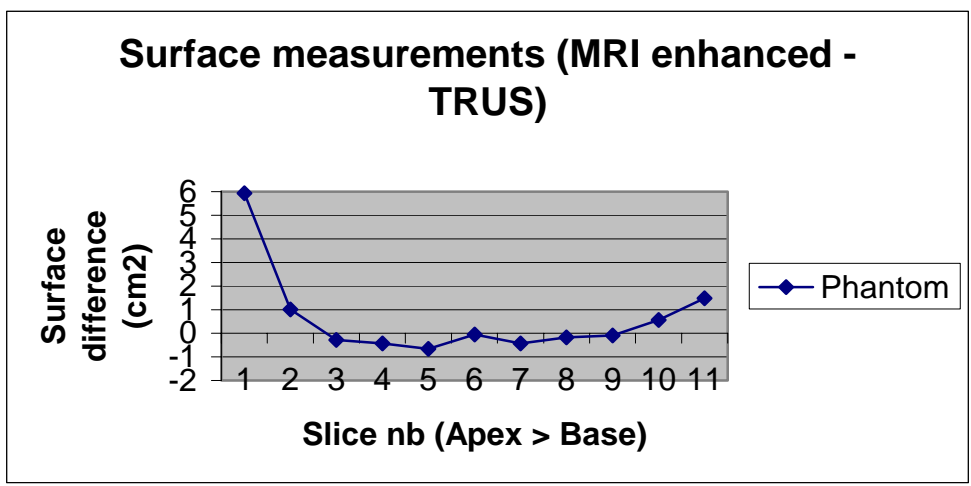

Figure 9 : *Surface analysis (phantom)* : elastic registration.

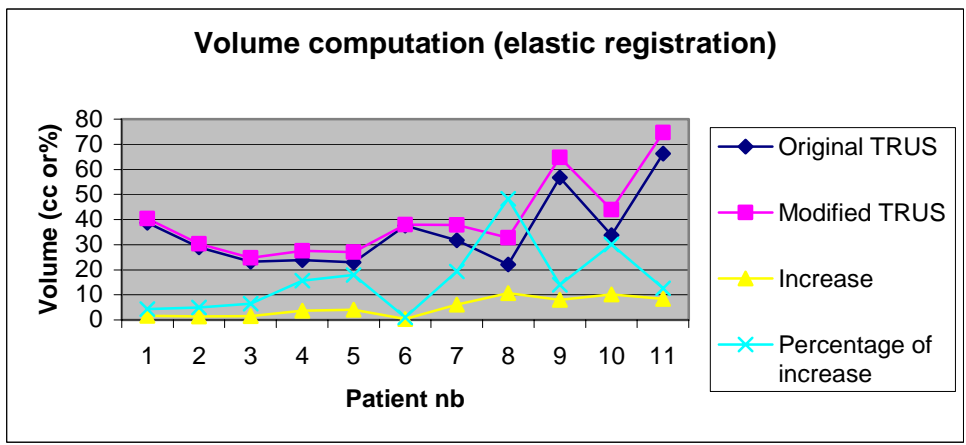

Figure 10: *Volume analysis (patients)*: elastic registration case